\documentclass[aps,twocolumn,showpacs]{revtex4}

\begin{document} 
\title{Periodic orbit quantization of the closed three-disk billiard
as an example of  a chaotic system with strong pruning}
\author{Kirsten Weibert, J\"org Main, and G\"unter Wunner}
\affiliation{Institut f\"ur Theoretische Physik 1,
         Universit\"at Stuttgart, D-70550 Stuttgart, Germany}
\date{\today}

\begin{abstract}
Classical  chaotic systems with  symbolic dynamics
but strong pruning present a particular
challenge for the application of
semiclassical quantization methods.
In the present study we show that the technique of
periodic orbit quantization by harmonic inversion of trace formulae, which 
does not rely on the
existence of a complete symbolic dynamics or other specific properties, 
lends itself ideally
to calculating semiclassical eigenvalues from periodic orbit data
even in strongly pruned systems. As  the number of 
periodic orbits proliferates exponentially in chaotic systems, 
we apply the harmonic inversion technique to cross-correlated
periodic orbit sums, which allows us to reduce the required number of orbits.
The power of the method is  demonstrated for the closed three-disk billiard
as a prime example of a classically chaotic bound system with strong pruning.
\end{abstract}


\maketitle

\section{Introduction}
The relation between quantum spectra, on the one hand, and the dynamics of the
corresponding classical systems, on the other, is a problem of fundamental 
importance in physics, which has attracted attention ever since the early 
days of quantum mechanics. 
A key to understanding this relation in chaotic systems was Gutzwiller's
discovery (cf.\ \cite{Gut90,Gut01}) that the semiclassical density of states 
in these systems can be written as a sum over all (isolated) periodic orbits 
of the  classical system. However, practical applications
of Gutzwiller's trace formula, like
other periodic orbit sums, are greatly impeded by the fact that the
sums usually diverge in the domain where the physical eigenvalues
are located, mainly as a consequence of  the rapid
proliferation of periodic orbits with growing period.
Various techniques were developed over the years to overcome the 
convergence problem of the periodic orbit sums
\cite{Cvi89,Aur92,Ber90},
but most of them depend on special properties of the systems, such as
ergodicity or the existence of  complete symbolic dynamics.

As an alternative, the technique of harmonic inversion of  semiclassical 
trace formulae has been introduced as a very general method
for extracting eigenvalues  from periodic orbit sums for both
chaotic and regular systems \cite{Mai97b,Mai98,Mai99a,Mai99b,Wei00,Mai00}. 
Harmonic inversion circumvents the convergence problems, and
allows one to obtain the
semiclassical eigenvalues from a relatively small number of periodic
orbits. Moreover, the method does not require any special properties 
of the systems,  and therefore is applicable also
to the particularly challenging case of  systems with strong
pruning, i.e., systems with a  highly incomplete  symbolic coding
of the periodic orbits. In this paper we demonstrate the power of the harmonic
inversion technique in the semiclassical quantization of the 
closed three-disk billiard system, for which  all other semiclassical 
quantization methods to date have failed in determining more than
the few lowest eigenvalues, chiefly because of the
extremely rapid proliferation of periodic orbits with increasing action
and the strong pruning typical of this system. From a technical point of 
view our study will show that harmonic inversion is numerically
stable even when handling huge (on the order of several million) 
periodic orbit sets.  

The three-disk billiard system  consists of three
equally spaced hard disks of unit radius.
The existence or nonexistence of periodic orbits and the 
behavior of the periodic orbit parameters in the three-disk
system sensitively depend on the distance $d$ between the centers
of the disks.
Large disk separations, 
especially $d=6$, have served as a test case  for periodic orbit quantization
of chaotic systems in many investigations in recent years.
In particular, the system has been employed as  prototype example for the 
usefulness of periodic orbit quantization by cycle expansion techniques
\cite{Cvi89,Eck93,Eck95,Wir99}.
For large disk separations, the assumptions of the cycle
expansion, namely that the contributions from long orbits are
shadowed by those of short orbits, are almost exactly fulfilled.
This is no longer true for small disk separations.
As the disks approach each other, the convergence of the cycle
expansion becomes slower and slower, until it finally breaks down.

Periodic orbit
quantization by harmonic inversion does not depend on specific properties,
such as  shadowing of orbits etc., and therefore is expected to 
work well also for small disk separations. In fact, 
the harmonic inversion technique has  successfully been 
applied to the ``standard'' literature disk separation $d=6$ as well
as to the small distance $d=2.5$, where the convergence of the
cycle expansion is already slow \cite{Mai98,Mai99a}.
In this paper we concentrate on the  limiting case of mutually 
touching disks, $d=2$, where the system turns into a bound system,
and the requirements of the cycle expansion
and other semiclassical methods are no longer fulfilled at all.

For the closed three-disk system, the lowest exact quantum eigenvalues were
first calculated by Tanner et al.~\cite{Tan91} and by Scherer \cite{Sche91}.
A remarkable step towards the
semiclassical quantization of the closed three-disk
system was achieved by combining the conventional cycle expansion with 
a functional equation \cite{Tan91}.
Based on periodic orbit corrections to the mean density of states,
approximations to the very lowest eigenvalues could be obtained from a small
set of periodic orbits up to cycle length $l=3$, where the symbolic dynamics
is still complete (including formally a zero length orbit).
For the very lowest eigenvalues, the results were found to be in good agreement
with the exact quantum values. 
However, the method failed for higher eigenvalues, where 
only the mean density of states could be reproduced.
Since the problems with the method arose from the strong pruning 
of orbits in this system, they are of a fundamental nature, and could not
be overcome by including more orbits.
No other semiclassical method has been successful in obtaining 
higher eigenvalues in the closed three-disk billiard
system yet. In this paper, we will demonstrate that harmonic inversion 
passes even this acid test of semiclassical quantization. 

In Sec.~\ref{3disk-po+qm} we 
discuss the construction of the periodic orbit signal for the closed 
three-disk system, which is a nontrivial task because of, firstly, 
the extremely rapid proliferation of periodic orbits and, secondly, the 
strong pruning of the symbolic dynamics.
Once the signal has been constructed, we refer in Sec.~\ref{HI} to
the harmonic inversion technique for the extraction of the semiclassical 
eigenvalues.
The numerical recipe of the harmonic inversion method has already been
established in the literature \cite{Mai97b,Mai98,Mai99a,Mai99b,Wei00,Mai00}, 
however, to make the article self-consistent we recapitulate the important 
steps in Sec.~\ref{method}.
The semiclassical results are discussed in Secs.~\ref{d=2_hi} and 
\ref{d=2_crosscorr}.

\section{Periodic orbits of the closed three-disk system }
\label{3disk-po+qm}
We start by summarizing the basic properties of the 
three-disk system relevant for periodic orbit quantization. 
Furthermore, the pruning of orbits and the distribution of the orbit 
parameters are investigated, and the resulting strategies for the 
numerical search for periodic orbits are discussed.

\subsection{Symbolic code and symmetry reduction}
The periodic orbits of the three-disk system can be labeled by a ternary
symbolic code, which is complete for sufficiently large disk separations.
The classical dynamics of the three-disk system has been
studied, among others, by Gaspard and Rice \cite{Gas89}.
If the disks are labeled by the numbers 1, 2, 3, each periodic orbit
is characterized by a sequence of these numbers,
indicating the disks the particle collides with during one period of the
orbit.
For sufficiently separated disks, there is a one-to-one correspondence 
between the symbolic code and the periodic orbits of the system:
For every sequence there exists one unique periodic orbit
(with the restriction that consecutive repetitions of the same symbol 
are forbidden and circular shifts of a sequence describe the same orbit).
As was first pointed out by Hansen  \cite{Han93}, at disk separation
$d=2.04821419$ ``pruning'' sets in, i.e., part of the periodic
orbits become unphysical as they begin to run through one of the disks, and 
thus the symbolic code is no longer complete.
The number of pruned orbits rapidly becomes larger and larger
if the disks continue to approach each other.
In the limiting case of touching disks, $d=2$, the system exhibits 
strong pruning.
Examples of pruned orbits will be discussed in Section \ref{numsearch}.

In our calculations, we  make use of the symmetry reduction introduced 
by Cvitanovi\'c and Eckhardt in Ref.~\cite{Cvi89}: 
The three-disk system is invariant under the symmetry
operations of the group $C_{3v}$, i.e., reflections at three symmetry
lines and rotations by $2\pi /3$ and $4\pi /3$.
The periodic orbits fall into three classes of distinct symmetry:
orbits invariant under reflections at one of the symmetry lines 
(multiplicity 3),
orbits invariant under rotations by $2\pi /3$ and $4\pi /3$ (multiplicity 2),
and orbits with no symmetry (multiplicity 6).
The quantum states are grouped in the three irreducible subspaces
$A_1$, $A_2$, and $E$, where the states of the $A_1$ ($A_2$)
subspace are symmetric (antisymmetric) under reflection at the symmetry
lines, respectively, and the states of the $E$ subspace are invariant
under rotations by $2\pi /3$ and $4\pi /3$.

\begin{figure}
\vspace{7.0cm}
\includegraphics{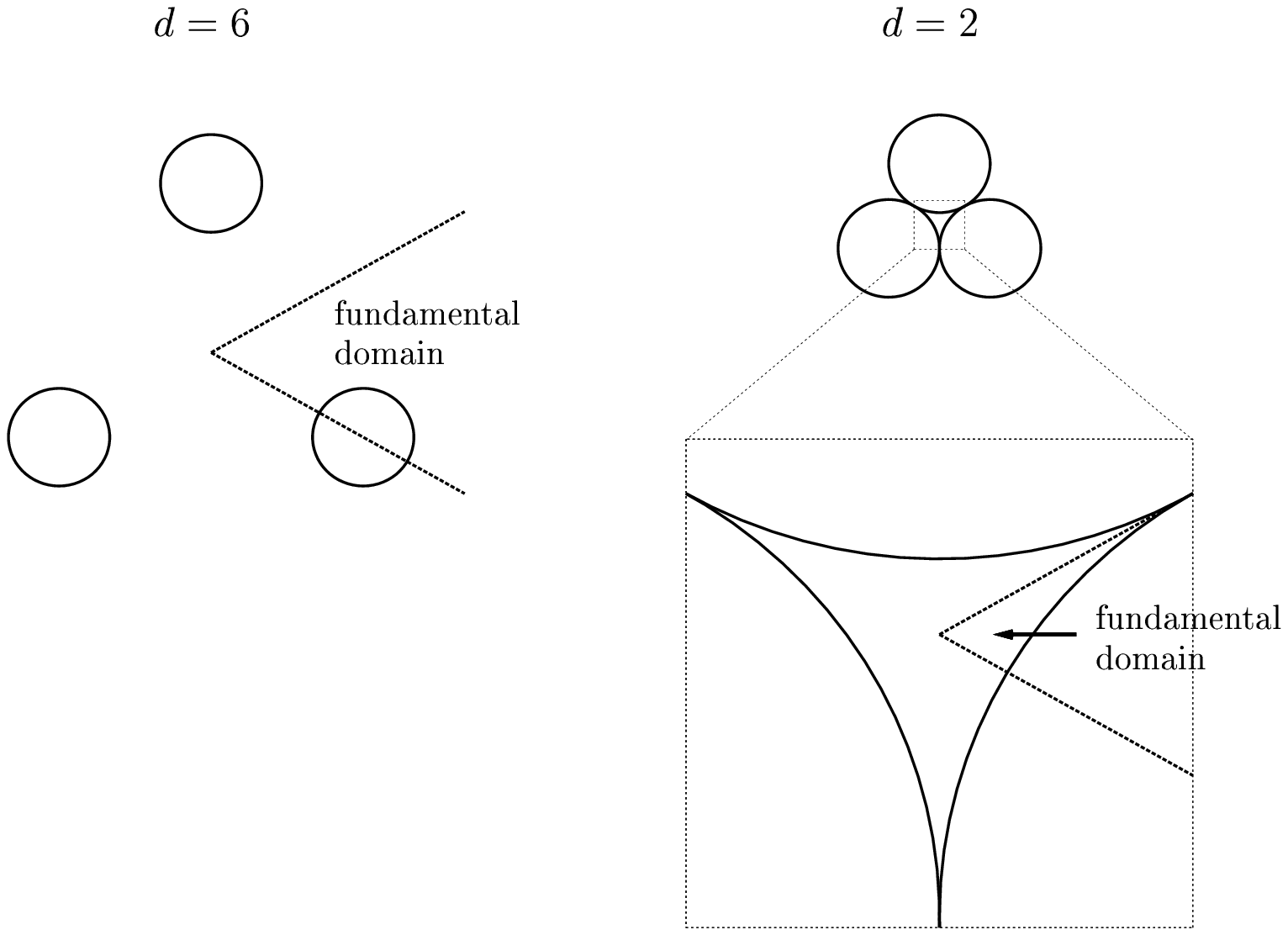}
\caption{
Open and closed three-disk system.
The fundamental domain consists of a one-sixth slice of the full system.
}
\label{fig1}
\end{figure}

Following Refs.~\cite{Cvi89,Eck95}, one can map the system onto a 
fundamental domain, which consists of a one-sixth slice of the full system, 
with the symmetry axes acting as straight mirror walls (see Fig.~\ref{fig1}).
The periodic orbits of the full system can  be  described completely
in terms of the periodic orbits in the fundamental domain.
The symmetry reduced periodic orbits are labeled by a binary code, 
where the symbol `0' represents backscattering 
(or change between  clockwise and anti-clockwise scattering)  
and the symbol `1' stands for scattering to the third disk 
in the original full domain picture.
For example,  the shortest full domain orbit 12 maps onto the 0 orbit,
and the 123 orbit maps onto the 1 orbit.

\subsection{Semiclassical density of states}
\label{3disk_dens_sec}
The quantum resonances of the three different subspaces $A_1$,
$A_2$ and $E$ can be obtained separately
from the periodic orbits in the fundamental domain by introducing 
appropriate weight factors for the orbits in Gutzwiller's trace formula,
as was shown by Cvitanovi\'c and Eckhardt \cite{Cvi93}.
We concentrate on the $A_1$ subspace, for which each orbit has a weight
factor equal to 1.

As for all billiard systems, the shape of the periodic orbits is independent 
of the wave number $k=\sqrt{2mE}/\hbar$, and the action scales as
\begin{equation}
S/\hbar= k s,
\end{equation}
where the scaled action $s$ is equal to the physical length of the orbit.
We consider the density of states 
as a function of the wave number
\begin{equation}
 \rho(k) = -{1\over\pi}\, {\rm Im}\ g(k) \; ,
\end{equation}
with a scaled response function $g(k)$.
Applying the Gutzwiller trace formula to the closed three-disk
system yields for the $A_1$ subspace
\begin{eqnarray}
\label{scGutz}
 && g(k) = g_0(k)
 -{\rm i}\sum_{\rm po} 
    {s_{\rm po}\, {\rm e}^{-{\rm i}{\pi\over 2}\mu_{\rm po}}
    \over r |\det(M_{\rm po}-1)|^{1/2}}\ 
    {\rm e}^{{\rm i}ks_{\rm po}} \\ \nonumber
&& = g_0(k) - {\rm i} \sum_{\rm po} (-1)^{l_s} 
     {s_{\rm po}\over r |(\lambda_{\rm po} -1)
     ({1\over \lambda_{\rm po}}-1)|^{1/2}}\ {\rm e}^{{\rm i}ks_{\rm po}} \, ,
\label{gw3disk}
\end{eqnarray}
where $M_{\rm po}$ and $\mu_{\rm po}$ are the monodromy matrix and the 
Maslov index of the orbit,
respectively,
$l_s$ is the symbol length, $s_{\rm po}$ is the scaled action,
 and $\lambda_{\rm po}$ denotes the expanding 
stability eigenvalue of the orbit
(i.e., the eigenvalue with an absolute value larger than one).
The sum runs over all symmetry reduced periodic orbits including multiple 
traversals. 
Here, $r$ denotes the repetition number with respect to the corresponding 
primitive orbit.

In practice, only the primitive periodic orbits  have to be determined.
The parameters of the $r^{\rm th}$ repetition of the primitive orbit 
(here characterized by the index 0) are then given by $l_s=rl_{s0}$,
$s=rs_0$ and $\lambda=\lambda_0^r$.

\subsection{Numerical search for periodic orbits}
\label{numsearch}
For extracting the quantum resonances of the system from Eq.~(\ref{gw3disk}) 
by harmonic inversion, all 
periodic orbits 
up to a maximum scaled action have to be included. 
The parameters of the periodic orbits -- scaled action and stability
eigenvalues -- have to be determined numerically.
We calculate the primitive periodic orbits 
using the symbolic code as input.
For simplicity, the calculations are carried out in the full domain,
and the results are then translated back into the symmetry reduced system.
The disks are ``connected'' according to the code, starting
with arbitrary reflection points on the disks as initial condition.
The reflection points are then varied in such a way that the total length 
(i.e., the action) of the orbit reaches a minimum.
All orbit parameters can then be calculated from the reflection
points. The scaled action is given by the length of the 
(symmetry reduced) orbit, and the 
stability eigenvalue $\lambda$ can be determined by an algorithm proposed
by Bogomolny \cite{Bog88}. 
In addition to action and stability, one can also determine 
averages of different classical quantities (distance from the
center of the system, angular momentum, etc.), which are needed for
the cross-correlation technique (see Section \ref{HI}).

\begin{figure}
\vspace{4.2cm}
\includegraphics{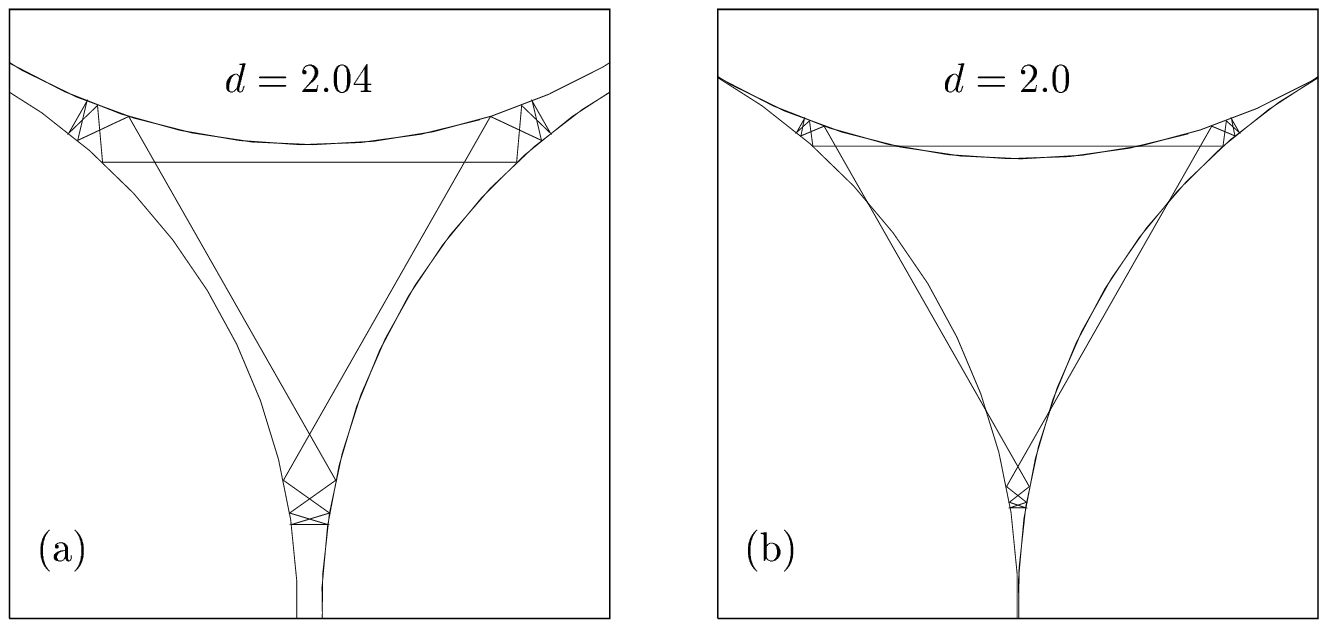}
\caption{
An example of the first class of pruned orbits (see text),
plotted in full domain representation. 
The symmetry reduced code of the orbit is 000011.
The orbit is shown at two different disk separations $d$, indicated
at the top of each diagram. 
At $d=2.04$, the orbit is still physical,
at $d=2.0$ it has become unphysical as is penetrates the disks.}
\label{fig2}
\end{figure}

\begin{figure}
\vspace{12.8cm}
\includegraphics{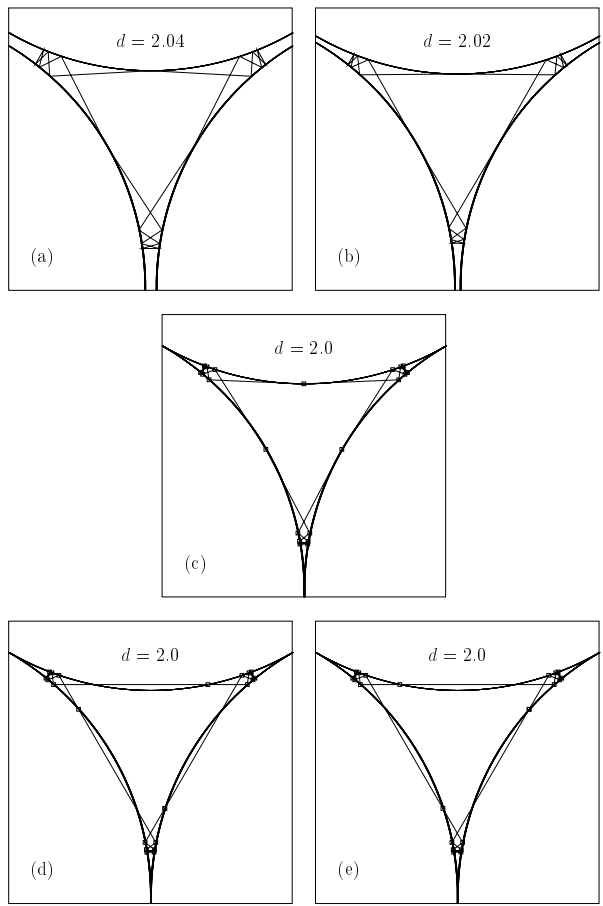}
\caption{
An example of the second class of pruned orbits (see text),
plotted in full domain representation. 
The symmetry reduced code of the orbit is 0000001.
The orbit is still physical at disk separations $d=2.04$ and $d=2.02$.
At $d=2.0$ it has split into three unphysical orbits. 
The squares indicate the ``reflection points''. Note in (c) that  the
reflection angle at the three central reflection points is larger
than $\pi$, i.e., the orbit in fact penetrates into the disks in the
vicinity of these  points.
}
\label{fig3}
\end{figure}

For disk separations smaller than the pruning limit $d=2.04821419$, it
has to be checked whether or not the orbits are physical, i.e., 
whether or not they 
stay completely outside the disks. In our numerical calculations, 
we found two different classes of pruned orbits.
An example of the first class is given in Figure \ref{fig2}:
As the disks approach each other, a section of the orbit connecting
two of the disks gets inside the third disk.
The pruned orbit still corresponds to a unique minimum of the
total length when the reflection points on the disks are varied for given
symbolic code.
This is not the case for the second class, 
an example of which is shown in Figure \ref{fig3}:
In this case,
as the disks come closer, the reflection angle at one specific reflection
point approaches $\pi$. 
As the value $\pi$ is reached, 
the minimum of the action with respect to this specific reflection point
splits into two equal minima and one local maximum.
Accordingly, the orbit splits into three unphysical 
orbits with the same symbolic code.

In one case, the orbit is now reflected on the inside of the disk
(see Fig.~\ref{fig3}c), which corresponds to a reflection angle
larger than $\pi$ (i.e., the orbit has penetrated into the disk
in the vicinity of  the reflection point).
This orbit does not correspond to a minimum of the
total length but to a saddle point.
In the other two cases, Figs.~\ref{fig3}d and \ref{fig3}e,
the reflection point moves along the disk in such a way that the
reflection law is no longer fulfilled but the orbit just passes
the ``reflection point'' in a straight line and penetrates into the disk.
These two cases correspond to two equally deep minima of the total
length. The two orbits have the same shape but only differ 
as to which
penetration point is considered as ``reflection point'', indicated by
the squares in Figs.~\ref{fig3}d and \ref{fig3}e.
In fact, the shape of the orbit is equal to that of a pruned orbit 
with a (symmetry reduced) symbol length that is shorter by 1 and
which belongs to the first class of pruned orbits as defined above.
(In this orbit the reflection point in question is simply missing,
i.e., none of the penetration points is considered as reflection point.)

\begin{figure}
\vspace{12.0cm}
\includegraphics{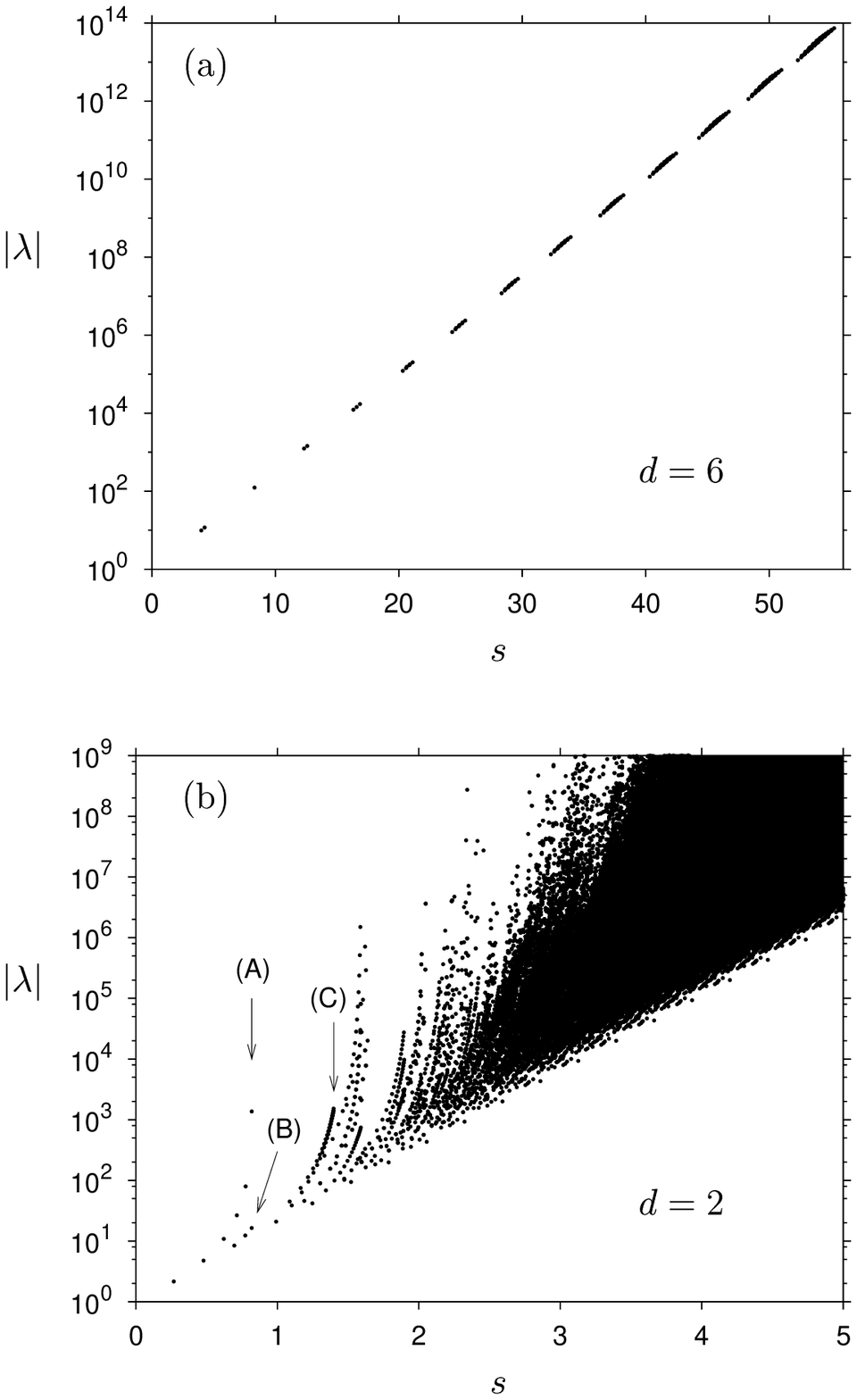}
\caption{
Distribution of periodic orbit parameters of the three-disk
system (a) with disk separation $d=6$, and (b) in the case of touching disks,
$d=2$.
The quantities plotted are the absolute value of the larger stability 
eigenvalue $|\lambda|$ versus the physical length $s$ of the orbits.
For $d=2$, for clarity a few channels of orbits are marked by arrows.
Channels (A) and (B) break off because of pruning.
The two channels marked (C) contain infinitely many orbits, but have 
been cut off by restricting the maximum number of successive zeros in 
the symbolic code (see text).
}
\label{fig4}
\end{figure}

Besides the existence or nonexistence of periodic orbits,
the distance between the disks also strongly influences the 
distribution of the periodic orbit parameters.
This is illustrated in 
Figure \ref{fig4}, which compares the orbit parameters
of the shortest 
primitive periodic orbits found for the open three-disk system with
disk separation $d=6$ and for the case of touching disks, $d=2$.
For $d=6$,
the action and stability of the orbits are mainly determined by the symbol 
length, the orbit parameters are nicely bunched in easily distinguishable
intervals, 
and the number of orbits up to a given action is relatively small.
However, the picture changes completely when the disks approach each other:
As all orbits become shorter, the total number of orbits up to a given 
action increases rapidly. The parameters of the orbits are no longer simply 
determined by the symbol length, and their overall behavior becomes much more 
complicated, as is evident from Figure \ref{fig4}.

\begin{figure}
\vspace{13.0cm}
\includegraphics{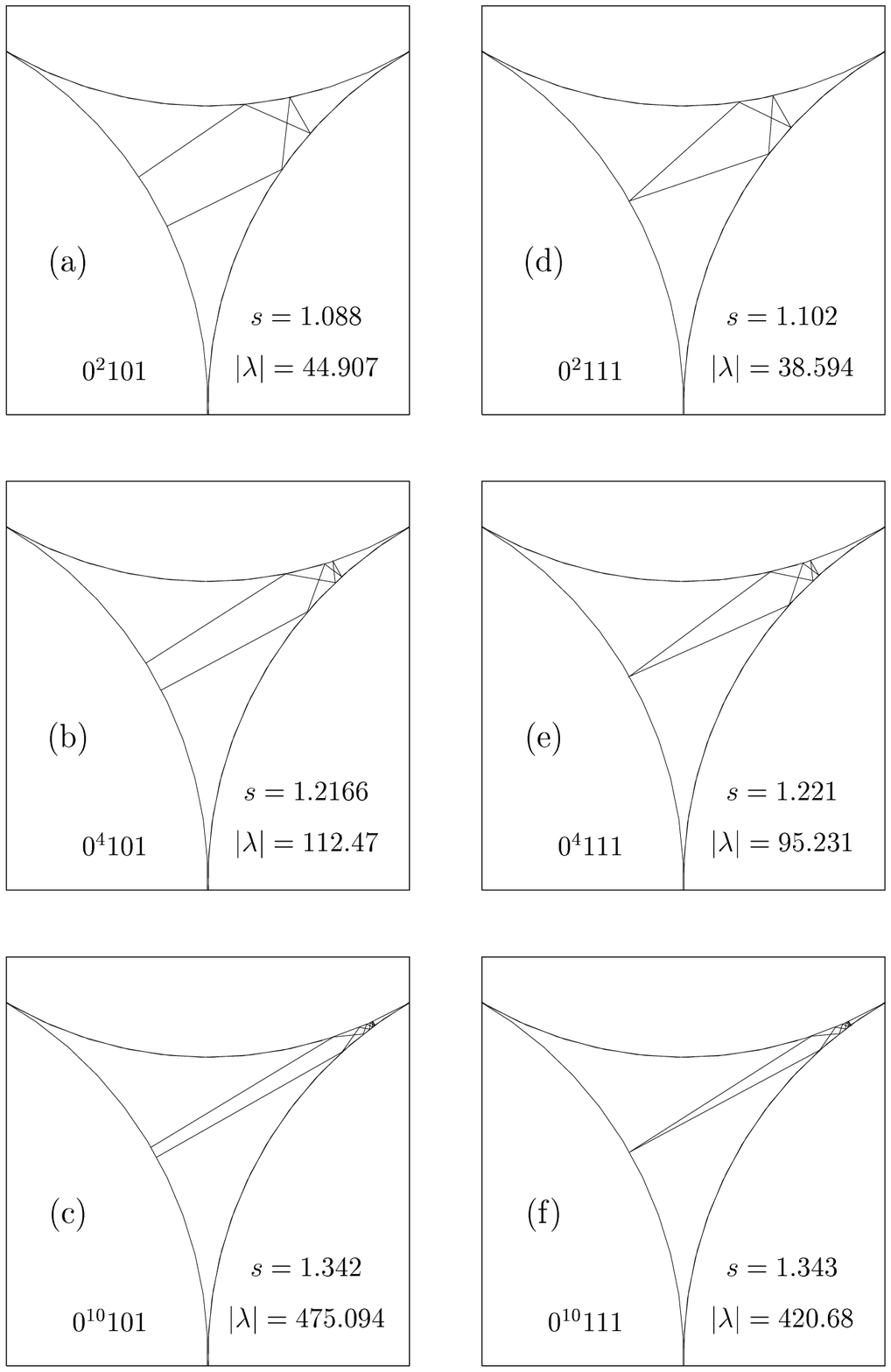}
\caption{
Examples of orbits of  the $0^n 101$ channel, (a) -- (c),  and 
the $0^n 111$ channel, (d) -- (f)  (dubbed (C) in Fig.~\ref{fig4}b), 
in full domain representation.
With growing number of leading `0's, the orbits run deeper and deeper into
the corners between the disks.
The series converges towards the limiting orbit that starts
exactly in the point where the disks touch and is then directly reflected back
from the opposite disk.
}
\label{fig5}
\end{figure}

For $d=2$, the orbits can be grouped in ``channels'' with the same ``tail'' 
(end figures) but growing number of leading `0's in the code.
(A sequence of $n$ leading `0's in the code will  be 
denoted by $0^n$ in the following).
These orbits have the same basic shape but run deeper and deeper into 
the corner formed by two touching disks, bouncing back and forth between 
the two disks (see Fig.~\ref{fig5}).
In each channel, the action of the orbits grows very slowly with
increasing symbol length, while the expanding stability 
eigenvalue increases relatively fast 
(in fact exponentially) with every additional collision.
Since adding a leading `0' to the symbolic code  does not 
change the action considerably,
there is a huge number of orbits with  very long symbol 
lengths but  relatively small, and comparable, actions.
In fact, most channels break off because of pruning. 
E.g., the $0^n 1$ series, labeled (A) in Fig.~\ref{fig4}b,
breaks off after six orbits, and the $0^n 11$ series,
marked (B), already breaks off after three orbits.
Our analysis strongly suggests that there exist only two infinite channels:
The $0^n 101$ and the $0^n 111$ series, dubbed (C) in Fig.~\ref{fig4}b
(these channels are only shown up to 20 leading `0's in the code).
The shapes and actions of these orbits converge very slowly
towards the shape and action of the limiting
orbit which starts exactly in the touching point of two disks and is
directly reflected back by the opposite disk (cf.~Fig.~\ref{fig5}).
Although all other channels break off because of pruning, the number 
of orbits in a channel up to a given action often becomes very large 
(or, in the two cases mentioned, even infinite).
Note that the 0 orbit, bouncing back and forth between two disks,
does not exist in the closed three-disk system.

While for large disk separations
finding all orbits up to a given action simply corresponds
to calculating all orbits up to a certain maximum symbol length,
the complicated distribution of periodic orbit parameters in 
the closed three-disk system renders the search for the 
relevant periodic orbits up to a given action a nontrivial task.
Due to the extremely rapid proliferation of periodic orbits with
increasing length $s$,
we were not able to determine all orbits up to a given length
$s_{\rm max}$, 
but we had to introduce restrictions for the orbits to be included.
As the amplitude in the Gutzwiller formula  depends on the inverse of the
stability parameter, a reasonable cut-off criterion is the size
of the stability of the orbit. 
On the other hand, since in each channel the absolute value of the expanding
stability eigenvalue grows exponentially fast,
the main contributions of each channel to the periodic orbit sum 
come from the orbits with small symbol lengths.
Furthermore, the Maslov indices of adjacent orbits in each channel differ by 
two, and therefore the periodic orbit amplitudes alternate in sign, which
leads to approximate cancellations of terms in the semiclassical trace
formula.
As a second criterion, we therefore restricted the number of consecutive 
symbols `0' in the symbolic code, thus selecting the most relevant orbits 
of each  channel.

The orbits were calculated channel by channel, starting from the shortest 
orbit in the channel and adding more and more leading zeros to the code.
The calculation of the channel was broken off if one of the following
conditions was fulfilled:
\begin{itemize}
\item The action exceeds the maximum action $s_{\rm max}$.
\item The absolute value of the stability eigenvalue exceeds the given value
      $\lambda_{\rm max}$.
\item The maximum number of consecutive symbols `0' is reached.
\item The orbits become pruned.
\end{itemize}
For the set of periodic orbits shown in Figure \ref{fig4}b,
we searched for orbits with physical length $s<s_{\rm max}=5.0$.
The maximum for the absolute value of the stability eigenvalue was chosen to be
$\lambda_{\rm max}=10^9$. The maximum number of 
successive zeros in the code was restricted to 20 for orbits with length 
$s<3.1$ and to 12 for orbits with $3.1<s<5.0$.

\section{Periodic orbit quantization of the closed three-disk system
by harmonic inversion}
\label{HI}
The fundamental problem, and challenge, now is to extract semiclassical 
eigenvalues in a numerically stable way from the huge periodic orbit set 
presented in Fig.~\ref{fig4}b.
To this end, we resort to harmonic inversion of periodic orbit signals
\cite{Mai97b,Mai98,Mai99a,Mai99b,Wei00,Mai00}.
Here, we review the basic ideas and present an extension of the method 
introduced in \cite{Mai00} to harmonic inversion of cross-correlated 
periodic orbit sums.

\subsection{The method}
\label{method}
The starting point is to introduce a weighted density of states in terms of $k$
\begin{equation}
   \varrho_{\alpha\alpha'}(k)
 = -{1\over\pi} \, {\rm Im} \, g_{\alpha\alpha'}(k) \; ,
\end{equation}
with
\begin{equation}
   g_{\alpha\alpha'}^{\rm qm}(k)
 = \sum_m {b_{\alpha n}b_{\alpha' n} \over k-k_m+{\rm i}\epsilon}\ ,
\label{g_ab_qm}
\end{equation}
where $k_m$ is the eigenvalue of the wave number of eigenstate $|m\rangle$
and
\begin{equation}
 b_{\alpha m} = \langle m|\hat A_\alpha|m\rangle
\end{equation}
are the diagonal matrix elements of a chosen set of $D$ linearly independent
operators $\hat A_\alpha$, $\alpha=1,2,\dots, D$.
The Fourier transform of (\ref{g_ab_qm}) yields a $D\times D$
cross-correlated signal
\begin{eqnarray}
 C_{\alpha\alpha'}^{\rm qm}(s) &=& {{\rm i}\over 2\pi}\int_{-\infty}^{+\infty}
      g_{\alpha\alpha'}(k){\rm e}^{-{\rm i}sk}{\rm d}k  \nonumber \\
 &=& \sum_m b_{\alpha m}b_{\alpha' m} {\rm e}^{-{\rm i}k_m s} \; .
\label{C_ab_qm}
\end{eqnarray}
A semiclassical approximation to the weighted quantum response function 
(\ref{g_ab_qm}) is given by the semiclassical response function 
(\ref{scGutz}) weighted with the classical averages
\begin{equation}
 a_{\alpha,{\rm po}} = {1\over s_{\rm po}} \int_0^{s_{\rm po}}
  A_\alpha({\bf q}(s),{\bf p}(s)) {\rm d}s \; ,
\label{a_po}
\end{equation}
with $A_\alpha({\bf q},{\bf p})$ the Wigner transform of the operator
$\hat A_\alpha$ \cite{Mai99c,Hor00}.
By Fourier transformation we obtain
the semiclassical approximation to the quantum signal
(\ref{C_ab_qm}), which reads
\begin{eqnarray}
   C_{\alpha\alpha'}^{\rm sc}(s)
 &=& \sum_{\rm po} {a_{\alpha,{\rm po}}\, a_{\alpha',{\rm po}}\,
    s_{\rm po}\, {\rm e}^{-{\rm i}{\pi\over 2}\mu_{\rm po}}
    \over r |\det(M_{\rm po}-{\bf 1})|^{1/2}}\
    \delta\left(s-s_{\rm po}\right) \nonumber \\
 &\equiv& \sum_{\rm po} {\cal A}_{\alpha\alpha'}^{\rm po}
    \delta\left(s-s_{\rm po}\right) \, ,
\label{C_ab_sc}
\end{eqnarray}
where $r$ is the repetition number counting the traversals of the primitive 
orbit, and $M_{\rm po}$ and $\mu_{\rm po}$ are the monodromy matrix and 
Maslov index of the orbit, respectively.
Semiclassical approximations to the eigenvalues $k_m$, and eventually also
to the diagonal matrix elements $\langle m|\hat A_\alpha|m\rangle$, are
obtained by adjusting the semiclassical cross-correlated periodic orbit
signal (\ref{C_ab_sc}) to the functional form of the quantum signal
(\ref{C_ab_qm}).
This can be done with the filter-diagonalization method for the signal 
processing of cross-correlation functions \cite{Nar97,Man98}.
Here, we resort to an alternative method introduced in Ref.~\cite{Mai00}, 
which will now be extended for the harmonic inversion of cross-correlated 
periodic orbit sums.

The special form of the periodic orbit signal (\ref{C_ab_sc}) as a 
sum of $\delta$ functions allows for an even simpler procedure than
filter-diagonalization, viz.\ analytical filtering.
In the following we will apply a rectangular filter, i.e., $f(k)=1$ 
for frequencies $k \in [k_0-\Delta k,k_0+\Delta k]$, and $f(k)=0$ outside 
the window.
Applying this filter to the semiclassical signal 
$C_{\alpha\alpha'}^{\rm sc}(s)$ in (\ref{C_ab_sc}) we obtain
the band-limited (bl) cross-correlated periodic orbit signal,
\begin{eqnarray}
     C_{\alpha\alpha'}^{\rm sc, bl}(s)
 &=& {{\rm i}\over 2\pi} \int_{k_0-\Delta k}^{k_0+\Delta k}
     g_{\alpha\alpha'}^{\rm sc}(k)
     {\rm e}^{-{\rm i}s(k-k_0)} {\rm d}k \nonumber \\
 &=& {{\rm i}\over 2\pi} \sum_{\rm po} {\cal A}_{\alpha\alpha'}^{\rm po}
     \int_{k_0-\Delta k}^{k_0+\Delta k}
     {\rm e}^{{\rm i}sk_0-{\rm i}(s-s_{\rm po})k} {\rm d}k \nonumber \\
 &=& {\rm i}\sum_{\rm po} {\cal A}_{\alpha\alpha'}^{\rm po}
     {\sin{[(s-s_{\rm po})\Delta k]}\over
     \pi(s-s_{\rm po})} {\rm e}^{{\rm i}s_{\rm po}k_0} \; .
\label{C_sc_bl}
\end{eqnarray}
The introduction of $k_0$ into the arguments of the exponential functions 
in (\ref{C_sc_bl}) causes a shift of frequencies by $-k_0$ in the frequency 
domain.
Note that $C_{\alpha\alpha'}^{\rm sc, bl}(s)$ is a smooth function and can 
be easily evaluated on an arbitrary grid of points $s_m<s_{\rm max}$ provided 
the periodic orbit data are known for the set of orbits with classical action
$s_{\rm po}<s_{\rm max}$.

Applying now the same filter as used for the semiclassical periodic orbit 
signal to the quantum one, we obtain the band-limited quantum signal
\begin{eqnarray}
 &&    C_{\alpha\alpha'}^{\rm qm, bl}(s)
 = {{\rm i}\over 2\pi} \int_{k_0-\Delta k}^{k_0+\Delta k}
     g_{\alpha\alpha'}^{\rm qm}(k)
     {\rm e}^{-{\rm i}s(ik-k_0)} {\rm d}k  \nonumber \\
 &=& \sum_{m=1}^M b_{\alpha m}b_{\alpha' m}
     {\rm e}^{-{\rm i}(k_m-k_0)s} \; ,    \; |k_m-k_0| < \Delta k \; .
\label{C_qm_bl}
\end{eqnarray}
In contrast to the signal $C_{\alpha\alpha'}^{\rm qm}(s)$ in 
Eq.~(\ref{C_ab_qm}), the band-limited quantum signal consists of a 
{\em finite} number of frequencies $k_m$, $m=1,\dots,M$, where in practical 
applications $M$ can be of the order of $\sim$~(50-200) for an appropriately 
chosen frequency window, $\Delta k$.
The problem of adjusting the band-limited semiclassical signal in
Eq.~(\ref{C_sc_bl}) to its quantum mechanical analogue in Eq.~(\ref{C_qm_bl})
can now be written as a set of nonlinear equations
\begin{eqnarray}
\label{C_bld}
 &&  C_{\alpha\alpha'}^{\rm sc, bl}(n\tau) \equiv c_{\alpha\alpha',n}
 = \sum_{m=1}^M b_{\alpha m}b_{\alpha' m}
   {\rm e}^{-{\rm i}k'_m n\tau} \; , \\ 
 && n = 0, 1, \dots, 2N-1 \; , \nonumber
\end{eqnarray}
for the unknown variables, viz.\ the shifted frequencies, 
$k'_m\equiv k_m-k_0$, and the parameters $b_{\alpha m}$, $m=1,\dots,M$.
Note that $M$ is related to the signal length $2N$ and the dimension $D$
of the cross-correlation matrix by $M=ND$.
The signal (\ref{C_bld}) can be evaluated on an equidistant grid, $s=n\tau$, 
with relatively large step size $\tau\equiv\pi/\Delta w$, and the discrete 
signal points 
$c_{\alpha\alpha',n}\equiv C_{\alpha\alpha'}^{\rm sc, bl}(n\tau)$ 
are called the ``band-limited'' cross-correlated periodic orbit signal.

We now wish to solve the nonlinear system, Eq.~(\ref{C_bld}), which can
be written as
\begin{equation}
 c_{\alpha\alpha',n} = \sum_{m=1}^M b_{\alpha m}b_{\alpha' m} z_m^n \; ,
\label{c_n:eq}
\end{equation}
where $z_m\equiv\exp{(-{\rm i}k'_m\tau)}$ and $b_{\alpha m}$ are, 
generally complex, variational parameters.
We assume that the number of frequencies in the signal is relatively small
($M\sim 50$ to $200$).
Although the system of nonlinear equations is, in general, still 
ill-conditioned, frequency filtering reduces the number of signal points, 
and hence the number of equations.
In Ref.~\cite{Mai00} three different methods, viz.\ linear predictor (LP), 
Pad\'e approximant (PA), and signal diagonalization (SD) have been employed
to solve Eq.~(\ref{c_n:eq}) for a one-dimensional signal, i.e., $D=1$.
The SD method can be generalized in a straightforward manner to the
harmonic inversion of cross-correlated periodic orbit sums.
The problem of solving the nonlinear set of equations (\ref{c_n:eq}) can 
be recast in the form of the generalized eigenvalue problem
\cite{Wal95,Man97,Man98}
\begin{equation}
 {\bf U} \mbox{\boldmath{$B$}}_m = z_m {\bf S} \mbox{\boldmath{$B$}}_m \; ,
\label{geneval}
\end{equation}
where the elements of the $M\times M$ operator matrix ${\bf U}$ and
overlap matrix ${\bf S}$ depend trivially upon the signal points:
\begin{eqnarray}
  U_{\alpha i,\alpha' j} &=& c_{\alpha\alpha',i+j+1} \; ; \nonumber \\
  \quad S_{\alpha i,\alpha' j} &=& c_{\alpha\alpha',i+j} \; ;
  \quad i,j=0,\dots,N-1 \; .
\label{matels}
\end{eqnarray}
The matrices ${\bf U}$ and ${\bf S}$ in Eq.~(\ref{geneval}) are complex
symmetric (i.e., non-Hermitian), and the eigenvectors 
$\mbox{\boldmath{$B$}}_m$ are orthogonal with respect to the overlap 
matrix ${\bf S}$,
\begin{equation}
   \left(\mbox{\boldmath{$B$}}_m |{\bf S}| \mbox{\boldmath{$B$}}_{m'} \right)
 = N_m \delta_{mm'} \; ,
\end{equation}
where the brackets define a complex symmetric inner product $(a|b)=(b|a)$,
i.e., no complex conjugation of either $a$ or $b.$
The overlap matrix ${\bf S}$ is not usually positive definite
and therefore the $N_m$'s are, in general complex, normalization parameters.
An eigenvector $\mbox{\boldmath{$B$}}_m$ cannot be normalized for $N_m=0$.
The parameters $b_{\alpha m}$ in Eq.~(\ref{c_n:eq}) are obtained from the 
eigenvectors $\mbox{\boldmath{$B$}}_m$ via
\begin{equation}
 b_{\alpha m} = {1\over\sqrt{N_m}}  \sum_{\alpha'=1}^D \sum_{n=0}^{N-1}
   c_{\alpha\alpha',n} \mbox{\boldmath{$B$}}_{m,\alpha' n} \; .
\end{equation}
The parameters $z_m$ in Eq.~(\ref{c_n:eq}) are given as the eigenvalues of 
the generalized eigenvalue problem (\ref{geneval}), and are simply related to 
the frequencies $k'_m$ in Eq.~(\ref{C_bld}) via $z_m=\exp(-{\rm i}k'_m\tau)$.

The resolution of the results depends on the signal length.
For a one-dimensional signal, the method requires a signal length 
of $s_{\rm max}\approx 4\pi\bar\rho(k)$, with $\bar\rho(k)$ the mean density
of states, to resolve the frequencies \cite{Mai97b}.
This means that all periodic orbits up to the scaled action $s_{\rm max}$ 
have to be included.
The advantage of using the cross-correlation approach is based on the 
insight that the total amount of independent information contained in
the $D\times D$ signal is $D(D+1)$ multiplied by the length of the signal, 
while the total number of unknowns (here $b_{\alpha m}$ and $k_m$) is $(D+1)$
times the total number of poles $k_m$.
Therefore the informational content of the $D\times D$ signal per unknown 
parameter is increased (compared to the one-dimensional signal) by 
roughly a factor of $D$, and the cross-correlation approach should lead to 
a significant improvement of the resolution.
The power of this method for periodic orbit quantization has already been 
proven for the example of an integrable system \cite{Mai99b,Wei00}.
We will now demonstrate how the method works in the case of 
three touching disks, as an example of a chaotic system with extremely
rapid proliferation of orbits and strong pruning.

\subsection{Semiclassical eigenvalues of the closed three-disk system by 
harmonic inversion of a one-dimensional signal}
\label{d=2_hi}
We first investigate a one-dimensional signal ($N=1$) by simply choosing
$\hat A_1={\bf 1}$, i.e., the unity operator.
The signal is constructed from the set of periodic orbits shown in 
Fig.~\ref{fig4}b, which contains about 5 million primitive orbits.

\begin{figure}
\vspace{5.8cm}
\includegraphics{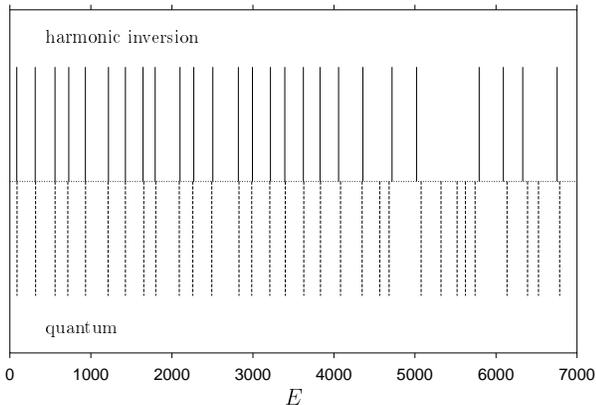}
\caption{
Energy eigenvalues of the closed three-disk system ($A_1$ subspace).
Dashed lines: exact quantum eigenvalues; solid lines: results 
$E=({\rm Re}\, k)^2/2$ from harmonic inversion of a signal of 
length $s_{\rm max}=4.9$.}
\label{fig6}
\end{figure}

We analyzed the signal up to length $s_{\rm max}=4.9$
using the decimated signal diagonalization method \cite{Mai00}.
The frequencies obtained by harmonic inversion of the signal are the 
semiclassical approximations to the eigenvalues $k_n$.
Fig.~\ref{fig6} shows the results from harmonic inversion (solid lines), 
compared with the exact quantum results (dashed lines), both presented 
in terms of the energy $E=({\rm Re}\, k)^2/2$.
The exact quantum eigenvalues 
were taken from  Scherer \cite{Sche91} and  Wirzba \cite{Wir}.
The harmonic inversion results clearly reproduce the quantum
eigenvalues up to $E\approx 4500$ with only small deviations.
However, the signal length was not sufficient to resolve the eigenvalues 
in the region $E>4500$, where the density of states with respect to the 
wave number $k$ (which grows $\sim\sqrt{E}$) becomes too large.

\begin{figure}
\vspace{6.5cm}
\includegraphics{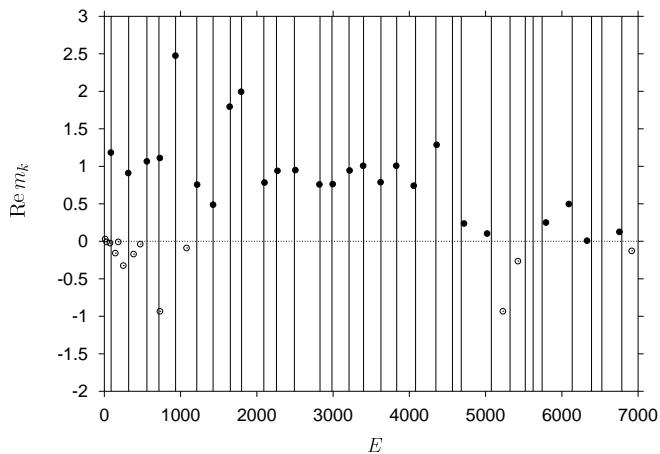}
\caption{
Amplitudes of the whole set of values $E=({\rm Re}\, k)^2/2$ (all circles)
obtained for the energy eigenvalues of the closed three-disk system by 
harmonic inversion (including also unconverged values).
The filled circles mark the values presented in Fig.~\ref{fig6}.
The solid vertical lines indicate the positions of the exact quantum 
eigenvalues.}
\label{fig7}
\end{figure}

Apart from the values shown in Fig.~\ref{fig6},
the harmonic inversion of the signal also yielded a number of
unconverged frequencies.
Converged frequencies were identified by having 
an imaginary part close to zero (since the eigenvalues of the wave number
must be real for bound systems) and an amplitude close to
the theoretical value $m_k=1$. 
The whole set of frequencies obtained is shown in Figure \ref{fig7} 
together with the real part of their amplitudes. The solid vertical lines 
mark the positions of the exact quantum eigenvalues.
The semiclassical values included in Fig.~\ref{fig6} 
are represented by filled circles in Fig.~\ref{fig7}.
The amplitudes are not as well converged as the frequencies
(i.e.\ the eigenvalues) but partly show larger deviations from 
the theoretical value $m_k=1$. 
[This can be explained by the observation that when the decimated signal 
diagonalization method is used for harmonic inversion the frequencies usually 
converge faster than the amplitudes.]
However, up to $E\approx 4500$, the eigenvalues can be clearly identified. 

The small deviations of the eigenvalues obtained by harmonic inversion 
from the exact quantum values may have different reasons.
Especially for the lowest eigenvalues, the major part is probably due to 
the semiclassical error.
On the other hand, as explained above, we did not include
all orbits in the signal but  left out a large
number of relatively unstable orbits. Although each of these orbits
only gives a negligible small contribution to the periodic orbit sum,
the number of excluded orbits may be so large that their
contributions sum up in a way so as
to have a visible effect on the density of states.
Finally, even in the low-lying part of the spectrum, deviations may arise
from the fact that the signal length was very short.
The influence of the missing orbits and the relatively short signal 
length are reflected in the relatively poor convergence of the amplitudes.

\begin{figure}
\vspace{6.5cm}
\includegraphics{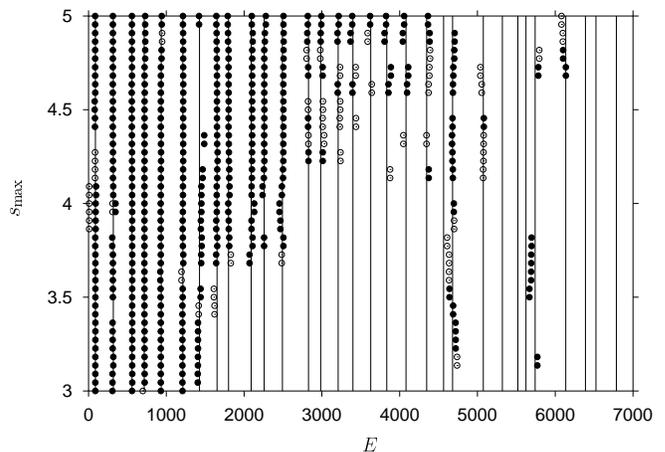}
\caption{
Harmonic inversion results $E=({\rm Re}\, k)^2/2$ for the energy eigenvalues of
the closed three-disk system as a function of the signal length $s_{\rm max}$.
Only results with $|{\rm Im}\, k|<1.0$ and amplitudes ${\rm Re}\, m_k>0.2$ 
are included.
The filled circles mark values with ${\rm Re}\, m_k>0.5$.
(Solid vertical lines: positions of the exact quantum energies.)}
\label{fig8}
\end{figure}

In order to test the stability of the results with respect to
the signal parameters, and to obtain an estimate of random errors,
we performed the same calculation with various sets of different signal
parameters.
Figure \ref{fig8} shows the harmonic inversion results for
the eigenvalues as a function of the signal length $s_{\rm max}$.
From all frequencies obtained the values with 
$|{\rm Im}\, k|<1.0$ 
and with  amplitudes 
${\rm Re}\, m_k>0.2$ were singled out, respectively. 
The filled circles mark amplitudes 
with ${\rm Re}\, m_k>0.5$. 
It can clearly be seen how the maximum eigenvalue up to which
the spectrum can be resolved depends on the signal length.
Fig.~\ref{fig8} shows that the results
for the low-lying eigenvalues are quite stable with respect to variation
of the signal length.

Compared with the results obtained by Tanner et al.~\cite{Tan91} using 
the extended cycle expansion method, 
the analysis of a single signal by harmonic inversion has 
yielded about the same number of eigenvalues for the closed three-disk system.
For the resolution of higher eigenvalues, the signal length was not sufficient.
Extending the signal to significantly larger scaled actions is in practice
not possible because of the extremely
rapid increase of the number of orbits.
It should be stressed, however, that, in contrast to the
cycle expansion method, which runs into severe problems because of 
the pruning of orbits in this system, it is at least {\it in principle} 
possible to improve the harmonic inversion results by including more orbits.
On the other hand, as discussed before,
the harmonic inversion method offers the possibility to significantly
reduce the signal length required for the resolution of eigenvalues 
by the construction and analysis of cross-correlated periodic orbit sums.

\subsection{Improvement of the resolution by harmonic inversion of 
cross-correlated periodic orbit sums}
\label{d=2_crosscorr}
To increase the accuracy of the semiclassical eigenvalues, we now
apply the cross-correlation technique to the closed three-disk system.
Instead of the one-dimensional signal investigated in the previous section,
we choose a set of three operators and construct a $3\times 3$ 
cross-correlated signal according to Eq.~(\ref{C_ab_sc}).
As input, we again use the set of periodic orbits presented in
Fig.~\ref{fig4}
together with the averages of different classical quantities over
the orbits.

\begin{figure}
\vspace{5.8cm}
\includegraphics{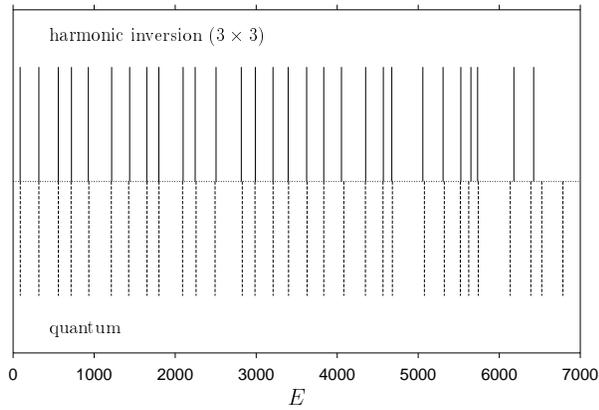}
\caption{
Energy eigenvalues of the closed three-disk system ($A_1$ subspace).
Dashed lines: exact quantum eigenvalues; solid lines: results 
$E=({\rm Re}\, k)^2/2$ from harmonic inversion of a $3\times 3$ 
cross-correlated signal with length $s_{\rm max}=4.8$. 
The operators chosen for the construction of the signal are ${\bf 1}$ (unity),
$r^4$ and $L^4$, where $r$ and $L$ are the distance from the center of the 
system and the absolute value of the angular momentum, respectively.}
\label{fig9}
\end{figure}

\begin{figure}
\vspace{13.0cm}
\includegraphics{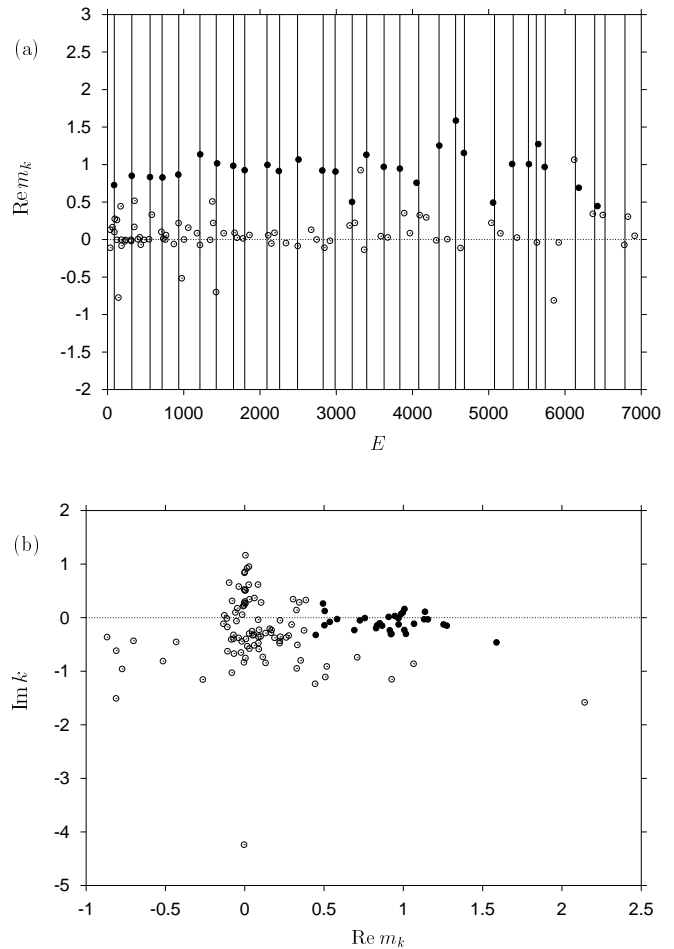}
\caption{
Closed three-disk system: The whole set of frequencies (all circles) resulting
from the analysis of the $3\times 3$ signal (see Fig.~\ref{fig9}). 
(a) Real part of the amplitudes versus energy $E=({\rm Re}\, k)^2/2$.
The solid vertical lines mark the positions of the exact quantum eigenvalues.
(b) Imaginary part of the frequencies versus real part of the amplitudes.
In both diagrams, the filled circles designate the values included in 
Fig.~\ref{fig9}.}
\label{fig10}
\end{figure}

Fig.~\ref{fig9} shows the results from the harmonic inversion of
a $3\times 3$ signal of length $s_{\rm max}=4.8$
(solid lines), which was constructed 
using the operators ${\bf 1}$ (unity), $r^4$ and $L^4$,
where $r$ and $L$ denote the distance from the center of the system and the
absolute value of the angular momentum, respectively. 
The eigenvalues are again presented in terms of the energy 
$E=({\rm Re}\, k)^2/2$.
For comparison,
the dashed lines indicate the positions of the exact quantum eigenvalues.
The converged eigenvalues for the wave number $k$ have been singled out 
from the whole set of frequencies obtained 
by the condition that they should have an imaginary part close to zero 
and an amplitude close to the theoretical value $m_k=1$. 
The amplitudes and the imaginary parts of all frequencies obtained
are shown in Figure \ref{fig10} by circles. 
In particular, the filled circles denote the frequencies included 
in Fig.~\ref{fig9}.
In Fig.~\ref{fig10}a, the positions of the exact quantum eigenvalues 
are given by the solid vertical lines.

With the cross-correlated signal, one can now clearly identify eigenvalues
up to $E\approx 6500$. 
In addition, the convergence of the lowest eigenvalues is improved as
compared to the results from the single signal obtained in Section
\ref{d=2_hi},
as can be seen from the amplitudes.
It is not possible to determine to what extent the accuracy of the
semiclassical eigenvalues has improved since the results can only
be compared with the exact quantum eigenvalues and the size of the 
semiclassical error is unknown.

\begin{figure}
\vspace{13.0cm}
\includegraphics{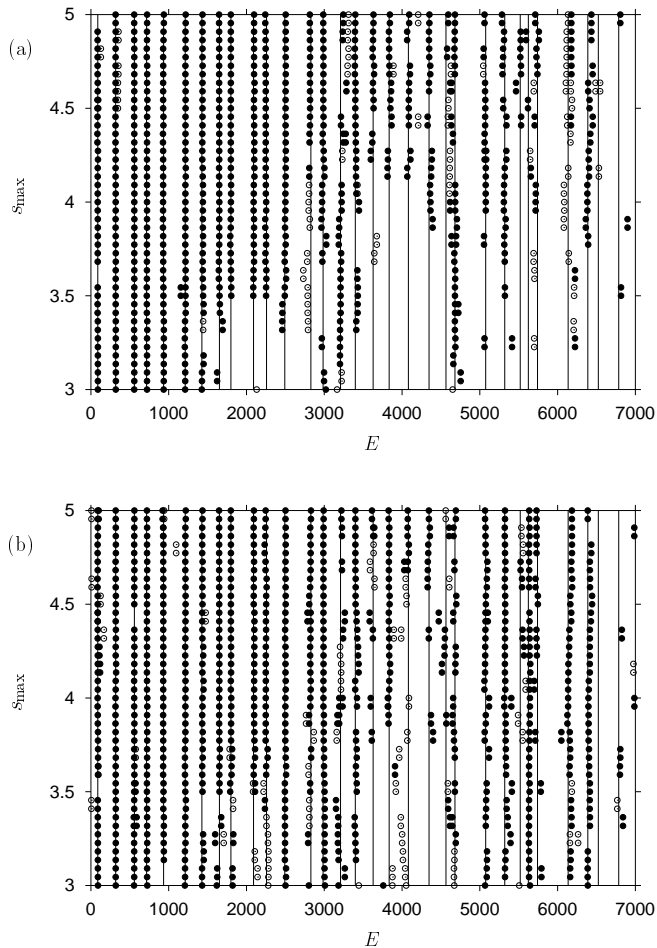}
\caption{
Harmonic inversion results $E=({\rm Re}\, k)^2/2$ for the energy eigenvalues 
of the closed three-disk system as a function of the signal length 
$s_{\rm max}$, calculated (a) from a $3\times 3$ cross-correlated signal 
including the operators ${\bf 1}$ (unity), $r^4$ and $L^4$ and (b) from a 
$4\times 4$ signal including the operators ${\bf 1}$, $r^2$, $r^4$ and $L^4$ 
($r$: distance from the center of the system, $L$: absolute values of the 
angular momentum).
Only results with $|{\rm Im}\, k|<1.0$ and amplitudes ${\rm Re}\, m_k>0.5$ 
are included.
The filled circles mark values with $|{\rm Im}\, k|<0.5$.
The solid vertical lines indicate the positions of the exact quantum 
eigenvalues.}
\label{fig11}
\end{figure}

Again, we have performed the same calculation for various sets of parameters
in the harmonic inversion scheme. Figure \ref{fig11} shows the 
results from a $3\times 3$ and from a $4\times 4$ signal as functions
of the signal length. The $3\times 3$ signal was constructed from the operators
${\bf 1}$ (unity), $r^4$ and $L^4$, and the $4\times 4$ signal contained
the operators  ${\bf 1}$, $r^2$, $r^4$ and $L^4$.
In both diagrams, only the frequencies
with an imaginary part $|{\rm Im}\, k|<1.0$ and an amplitude 
${\rm Re}\, m_k>0.5$ were included. 
The filled circles mark values with $|{\rm Im}\, k|<0.5$. 
The positions of the exact quantum eigenvalues are again marked by 
the solid vertical lines.

As with the one-dimensional signal analyzed in the previous section, 
it can be observed that the results for the lowest eigenvalues
are very stable with respect to the variation of the signal length
as long as the signal is not too short.
However, the very lowest frequencies show a tendency to split into two 
or more, which was a frequent observation in our calculations
when the signal length and the
matrix dimensions were chosen too large.
The results for the higher eigenvalues depend more sensitively on the
signal length as the density of states approaches the limiting resolution
that can be achieved with the signal.
In this region, the $4\times 4$ signal shows a better resolution than the
$3\times 3$ signal.

With the $3\times 3$ signal and the two largest values of the signal length 
considered,
the lowest eigenvalue was not obtained. 
The reason for this probably lies in inaccuracies at the end of the signal,
as the signal length approaches the value $s_{\rm max}=5.0$:
In the decimated signal diagonalization scheme, which we used for the harmonic
inversion of the signal, the $\delta$ functions in the original signal
(\ref{C_ab_sc}) are replaced with sinc functions \cite{Mai00}
[${\rm sinc}~ x = (\sin x)/x$].
Therefore, every periodic orbit contributes to the signal 
not only exactly at its scaled action $s=s_{\rm po}$, but also 
in a range $\Delta s$ around this point.
If the signal length is close to $s_{\rm max}=5.0$, also orbits 
with actions slightly larger than this limit may give non-negligible
contributions to the very
end of the signal. Since we did only include orbits with $s\le 5.0$,
these contributions are missing in our signal.

Some eigenvalues between $E=3500$ and $E=5000$ seem to be particularly
hard to obtain from the set of orbits used. 
Possibly, these eigenvalues are related to the orbits running very deep 
into the corners between the disks, which were not included in the signal. 
We conjecture that the missing orbits as well as the short signal length 
are again responsible for the deviations especially of the 
results for higher eigenvalues from the exact quantum values.

\section{Conclusion}
We have successfully applied the technique of harmonic inversion
to the  semiclassical
quantization of a system with an extremely rapid
proliferation of periodic orbits and strong pruning.
With the harmonic inversion of cross-correlated periodic
orbit sums we were able to calculate accurate semiclassical energy eigenvalues of the
closed three-disk system up to the region $E\approx 6500$ (a total of
29 levels).  This is the first time that such high-lying eigenvalues 
 have been correctly calculated by periodic orbit theory in this system.
Our results demonstrate that the harmonic inversion of cross-correlated
periodic orbit sums is indeed a powerful
method of periodic orbit quantization and
allows  the stable handling even of huge periodic orbit sets.
In contrast to other methods, which depend on the existence
of a complete symbolic code, the harmonic inversion 
technique is not affected by the strong pruning of orbits. 
In the closed three-disk system,
the only restriction for the practical application 
of the harmonic inversion method lies in the extremely
large number of periodic orbits, which is a special feature of
this system. 
It should also be noted that the semiclassical eigenvalues were
obtained without resorting to the mean density of states, which, including
all necessary correction terms, is nontrivial to calculate even in the case
of the closed three-disk system \cite{Sche91}.
The present results may stimulate future work on other challenging systems
without special structural information on the quantum or classical level.
One example with strong pruning of the symbolic dynamics is the 
three-dimensional generalization of the closed three-disk billiard, i.e., 
a system consisting of four touching spheres at the corners of a regular 
tetrahedron.
By contrast to the closed three-disk system this is an open system where
no functional equation can be applied.
Experiments on chaotic light-scattering from the four spheres have recently
attracted much attention \cite{Swe99}.

\acknowledgments
We thank A.\ Wirzba for communicating to us quantum mechanical data
for the closed three-disk system.
This work was supported by  Deutsche For\-schungs\-ge\-mein\-schaft and
Deutscher Akademischer Austauschdienst.


\end{document}